\documentclass[twocolumn]{aastex631}

\newcommand\new[1]{#1}
\newcommand\newer[1]{#1}
\newcommand\old[1]{}

\submitjournal{ApJ\old{L}}

\shorttitle{Square Root Compression \& Noise}
\shortauthors{DeForest et al.}

\graphicspath{{./}{}}

\begin{document}

\title{Square Root Compression and Noise Effects in Digitally Transformed Images}

\author[0000-0002-7164-2786]{Craig E. DeForest}
\affiliation{Southwest Research Institute \\
1050 Walnut Street, Suite 300 \\
Boulder, CO 80302, USA}

\author[0000-0001-8318-8229]{Chris Lowder}
\affiliation{Southwest Research Institute \\
1050 Walnut Street, Suite 300 \\
Boulder, CO 80302, USA}

\author[0000-0002-0494-2025]{Daniel B. Seaton}
\affiliation{Southwest Research Institute \\
1050 Walnut Street, Suite 300 \\
Boulder, CO 80302, USA}

\author[0000-0002-0631-2393]{Matthew J. West}
\affiliation{Southwest Research Institute \\
1050 Walnut Street, Suite 300 \\
Boulder, CO 80302, USA}

\begin{abstract}

We report on a particular example of noise and data representation
interacting to introduce systematic error.
\new{Many instruments collect integer digitized values and 
apply nonlinear coding, 
in particular square-root 
coding, to compress the data for transfer or downlink; this can introduce surprising
systematic errors when they are decoded for analysis.
\newer{Square root coding and subsequent decoding typically introduces a variable, $\pm 1$ count value-dependent systematic bias in the data after reconstitution.  This is significant when large numbers of measurements (e.g., image pixels) are averaged together.}  
Using direct modeling of the probability distribution of particular 
coded values in the presence of instrument noise, one may apply Bayes' 
Theorem to construct a 
decoding table that reduces
this error source to a very small fraction of a digitizer step\newer{; in our example, systematic error from square root coding is reduced by a factor of 20 from 0.23 count RMS to 0.013 count RMS}. The method is suitable both for 
new experiments such as the upcoming PUNCH mission,
and also for post facto
application to existing data sets -- even if the instrument noise properties are only 
loosely known. Further, the method does not depend on the specifics of the coding formula,
and may be applied to other forms of nonlinear coding or representation of data values.}

\end{abstract}


\section{Introduction} \label{sec:intro}

Scientific measurement generally includes both ``noise'', which is frequently treated as a zero-mean 
normally distributed 
random variable, uncorrelated across measurements; and systematic ``error'', which is typically 
correlated across samples, 
and/or has a nonzero mean value.  Systematic error is typically minimized by instrument
calibration, and would ideally be zero in a perfectly calibrated instrument.
Uncorrelated normally-distributed noise therefore drives the sensitivity of most
astronomical and heliophysical remote sensing: essentially every telescopic or
spectroscopic measurement includes Poisson\new{-distributed} noise associated with photon counting in each sample; 
and the Poisson distribution is very well approximated by a normal distribution for 
large numbers of photons \citep[e.g.,][]{Feller1971}.  This property is important because many
post-processing techniques, from commonly-used averaging or smoothing to more sophisticated
Fourier-domain \citep[e.g.,][]{Yaroslavsky1996,deBoer1996,DeForest2017} or AI-based 
\citep[e.g.,][]{ParkEtal2020} methods can reduce noise effects well below the nominal noise floor of a 
single measurement, provided that the noise itself is well behaved in the data products being 
processed.  

Under linear transformations of data values, including data vectors comprising 
several independent measurements,
normally distributed (``Gaussian'') noise 
is indeed well behaved. Its properties as a random variable may be treated with conventional 
rules of thumb, such as addition in quadrature to combine
multiple noise sources and/or ``beat down'' noise through averaging.  
However, under nonlinear transformation, normally
distributed noise is not in general well behaved, and can transform to different 
statistical distributions or even develop a nonzero mean value (becoming a source
of systematic error), depending on the specific transformation.  
A salient example is the recent insightful analysis by \citet{Inhester_etal2021}, 
in which a common angle-free
derived measure of polarized brightness in solar coronal measurements 
($^\circ pB \equiv \sqrt{Q^2+U^2}$ for Stokes parameters $Q$ and $U$)  
is shown to respond counter-intuitively to noise because the measure is nonlinearly
related to the primary brightness values and/or Stokes parameters that are
used in its construction.

In developing the Polarimeter to UNify the Corona and Heliosphere
\citep[PUNCH;][]{deforest2018} we
sought to reduce data volume being downlinked from orbit, by
square-root coding the data.  Square root coding reduces downlink
volume by reducing the number 
of bits in photometric data, without significant loss; it does this by matching
the digital transition size to the corresponding photon-counting 
noise level across the dynamic range of the instrument \citep[e.g.,][]{Gowen_Smith2003}.  It has been
used or considered for many data-constrained instruments on the ground and in space, including
Yohkoh/SXT \citep{Acton_etal1992}, SOHO/MDI \citep{Scherrer_etal1995}, 
GONG \citep{Goodrich_etal2004}, Hinode/XRT \citep{Golub_etal2007}, 
PROBA2/SWAP \citep[which uses a custom recoding function similar to the square root;][]{Seaton2013}, 
and JWST \citep[nee NGST;][]{Nieto_etal1999}.  The PUNCH application pushes the limits of the 
technique, because PUNCH specifically requires averaging many individual, independent brightness
measurements to \new{meet} its driving science \new{requirements for photometric sensitivity}.

As a polarimetric mission to study the extended solar corona against the much brighter
zodiacal light and background stars, PUNCH has quite stringent requirements of order 
$10^{-4}$ precision in relative photometric sensitivity across patches of image.  This level 
of precision requires averaging photometric values across many pixels in both space and time, 
and therefore requires reducing any systematic errors well below the noise level in any one pixel.   

In this article, we describe: the PUNCH square root coding system (Section \ref{sec:sqrt}); a systematic 
error intrinsic to direct arithmetic square root coding/decoding (Section \ref{sec:results}); 
and a better
square root decoder produced via Bayesian analysis (Section \ref{sec:better}). 
We close by discussing  
the results, \new{relating them to other prior work \citep[e.g.,][]{Bernstein_etal2010}} and generalizing to other instruments \new{and applications -- including post-facto improvement
of existing data from prior instruments or observations} (Section \ref{sec:DandC}).

\begin{figure*}
    \centering
    \includegraphics[width=6.5in]{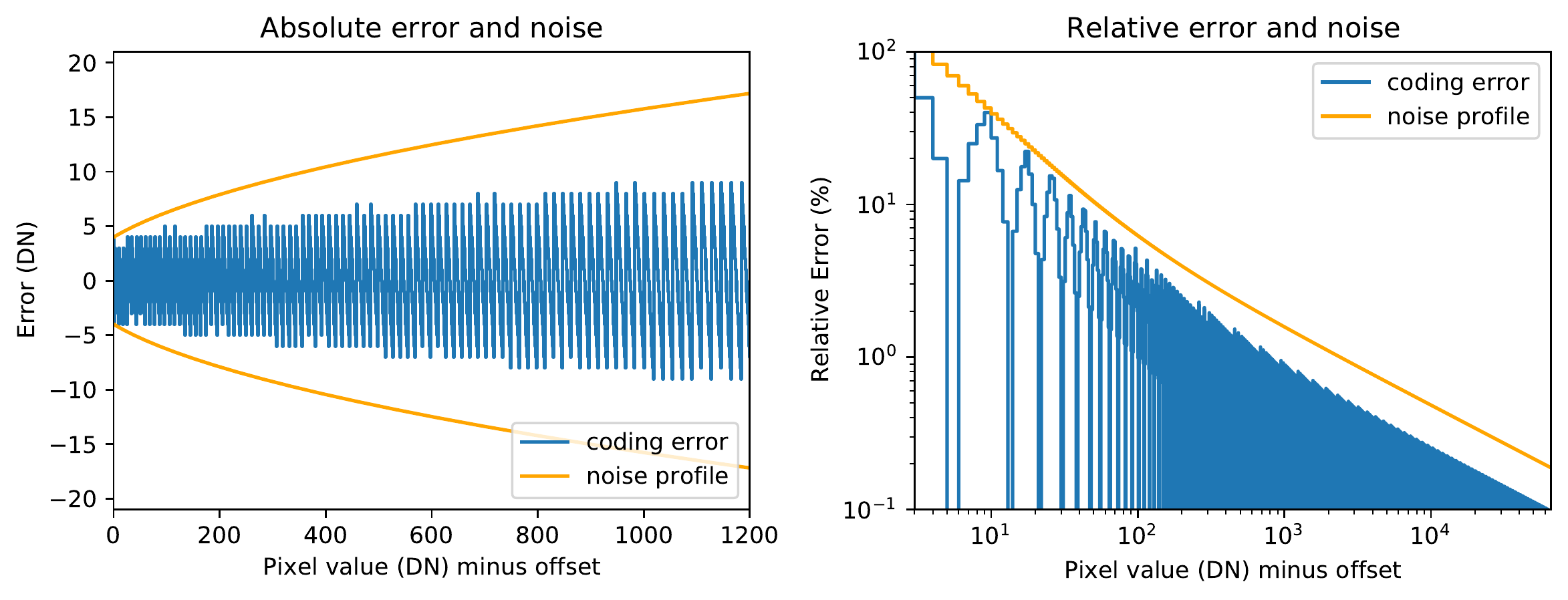}
    \caption{Square root coding is lossy and therefore induces value-specific error when decoded. Properly tuned coding schemes keep the coding error slightly below the other existing noise sources, saving bits and reducing unnecessary signal entropy.  Coding 16 bits to 10 bits as shown maintains the coding error a factor of 2-3 under the calculated existing noise level, throughout the dynamic range, for a particular noise model that includes both Poisson statistics and camera dark/read noise (see text).}
    \label{fig:coding}
\end{figure*}

\section{\label{sec:sqrt} Square Root Coding}

Square root coding is a simple lossy digital compression scheme that works by
matching the digital transition scale to anticipated photon noise level across the
dynamic range of an instrument \citep{Gowen_Smith2003}.  Taking the direct square root
of an (unsigned integer) data value reduces the number of bits by a factor of two.
In a system with one quantum (e.g., photon) count per digitizer number (DN), and no
detector offset, this matches the transition step size to the value-dependent
variance of the \new{quantum (``shot'')} noise \new{in each measurement}.  \citet{Gowen_Smith2003} discuss the development of a modified
square-root law that takes offset values into account: both digitizer gain and digitizer offset
in a typical detector.  PUNCH has a digitizer gain of 1 DN per 
4.3 detected photons, dark/read noise that is modeled as normally distributed with $\sigma=5$ DN, and a programmable camera offset intended to operate near 100-200 counts out 
of a digitizer range of $2^{16}$ counts.  We adopted the in-flight coding scheme
\begin{equation}
    \label{eq:coding}
    \new{c} = \sqrt{ a P }\,,
\end{equation}
where integer (rounded) arithmetic is assumed at each step (because it takes place onboard the 
spacecraft), $P$ is the pixel value
direct from the camera or summed across multiple exposures up to 19 bits, and \new{$c$} is the
coded pixel value.  The constant $a$ is programmable to generate  
values of \new{$c$} with between 9-14 bits of dynamic range.  In typical single-exposure operation,
16-bit camera values are square-root coded to 10-bit depth; \new{this is accomplished by setting $a=16$}.  
The values are then decoded on the ground to recover values approximating $P$.  We began with the simple integer-arithmetic decoding scheme
\begin{equation}
    \label{eq:decoding}
    B = ( c \times c ) \,/\, a\,,
\end{equation}
where the division is rounded to the nearest integer.

Figure~\ref{fig:coding} shows the distribution of quantization error, vs. Poisson
noise, over a typical dynamic range for the PUNCH camera.  The quantization error 
is always well inside the Poisson noise envelope, so that quantization is
dithered\footnote{Readers are reminded that dithering is the act of applying random or pseudo-random 
noise to an analog
signal before digitizing, to mitigate the effects of quantization error and related 
artifacts
on the analog signal.} 
by 1-4 steps across the entire dynamic range, retaining smooth representation of the
dynamic range, compared to the square root step size, when averaged across pixels. 
Information about each particular sample of the
Poisson noise is lost, reducing the entropy of the remaining bits and improving
\new{the compression ratio} for any subsequent compression.

\section{\label{sec:results} Initial results}

\new{
    \begin{figure*}
    \centering
    \includegraphics[width=6in]{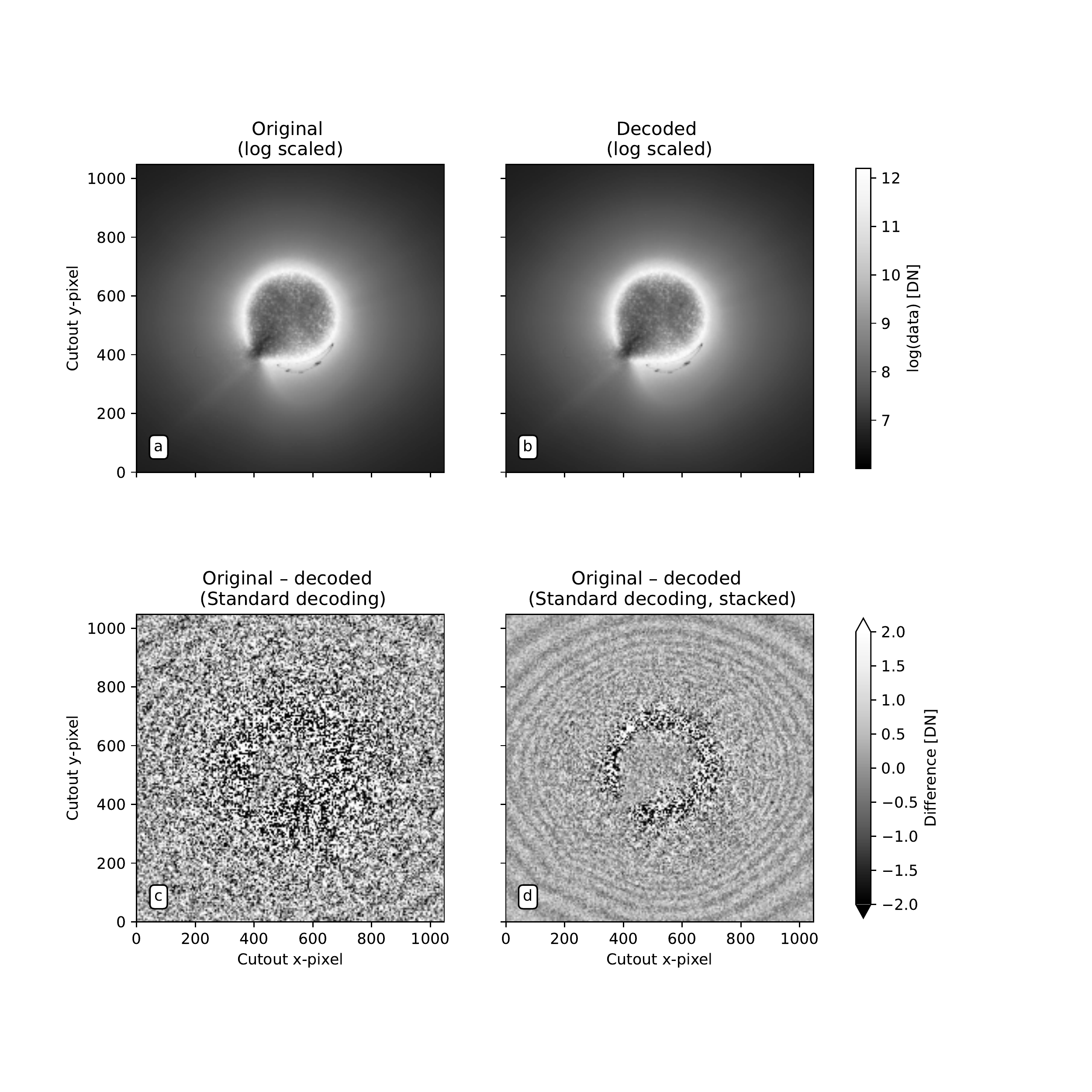}
    \caption{A visual comparison of an original dataset (a) and the encoded / decoded version (b) yields no immediately perceptible difference. However, when taking a difference map (c) and scaling to ±2 DN, a visual banding pattern appears. Beating down the noise by averaging 10 copies of the image, each with a different noise field, reduces the noise and reveals the systematic banding pattern (d).}
    \label{fig:diffmaps_original}
\end{figure*}
}

\begin{figure*}
    \centering
    \includegraphics[width=6in]{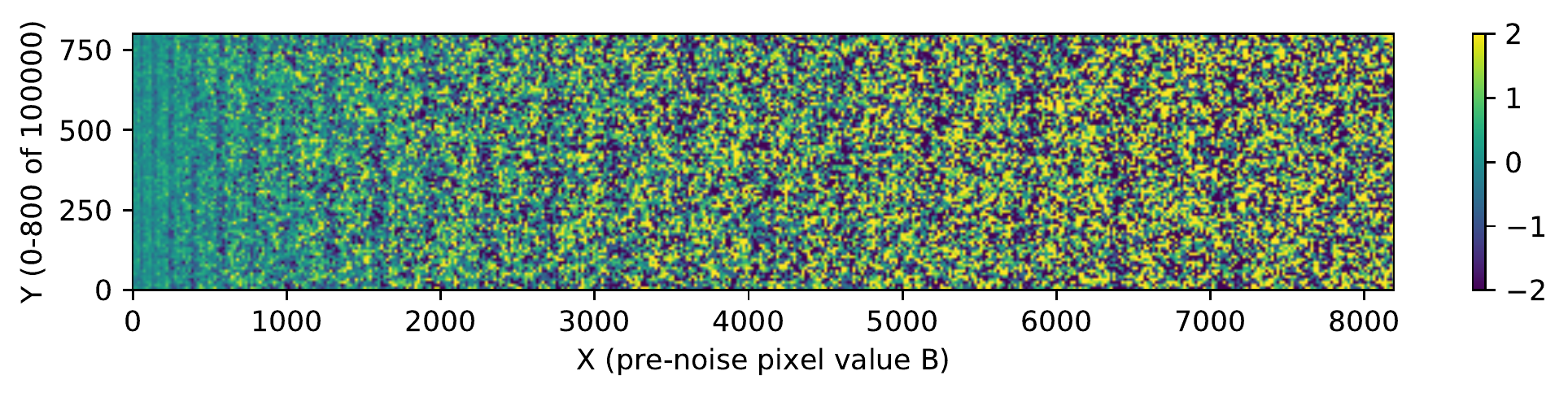}
    \caption{Difference between original and noisy, square-root coded versions of a simple gradient image shows
    vertical artifacts.  \new{The original image is 8192 pixels wide, and the value of each pixel is the X
    coordinate.  The image was degraded with modeled photon noise based on the CCD model in the text, then
    square-root coded from 16 to 10 bits and decoded again.  The figure shows the difference between the 
    processed image and the original gradient image.  Vertical stripes occur at values that the decoding
    systematically over- or under-estimates, revealing the source of the artifacts in 
    Figure~\ref{fig:diffmaps_original}(c-d).}}
    \label{fig:gradient}
\end{figure*}

We tested the encoding scheme in Section \ref{sec:sqrt} against existing data from
the STEREO/COR2 instrument \citep{Howard_etal2008}.  
We summed 32 images of floating-point COR2 data from a deep-field campaign
\citep{DeForest_etal2018}, scaled the result to PUNCH-like 16 bit values, 
added normally-distributed photon noise to match the PUNCH single-exposure
characteristics, square-root
coded the data, and then decoded them. Figure~\ref{fig:diffmaps_original} shows the 
original data \new{in panel} (a) and a reconstituted version, \new{coded with 
Equation ~\ref{eq:coding} and decoded with Equation~\ref{eq:decoding}, in panel} (b).
A difference image between the original (a) and reconstituted (b) images is also displayed in \new{panel} (c),
which is is scaled to a dynamic range of $\pm$2~DN.  Despite the fact that the digital transitions
are below the constructed noise floor of the dynamic range, square root coding produced 
visible artifacts in the smooth gradient of the F corona.  This is a residual effect that 
is present under correct transition levels that are fully dithered by the Poisson noise, and
is distinct from artifacts that are to be expected when the coded digital 
steps are comparable to (or larger than) the Poisson noise variance at a given pixel value.  The
peak amplitude of the artifacts can be seen in Figure \ref{fig:gradient} and is approximately
$\pm$0.5~DN, 
much smaller than the quantization errors
would be in the absence of dithering. (Quantization error on any one sample is roughly $\sqrt{P}/4$, or 10-80 DN in this dynamic range). 
To further illustrate the artifacts, we generated 10 copies of the source image, each with
an individual sample of the noise field. We averaged all the copies together to ``beat down'' the
noise 
after decoding, revealing \new{more clearly} the banding artifacts in \new{panel} (d).

To understand the artifacts in Figure~\ref{fig:diffmaps_original}, we analyzed a simple
horizontal gradient image whose value (in DN) was equal to the $x$ coordinate (in pixels),
over a horizontal range of 0--8192 (0--$2^{13}$), with 100,000 samples of each input value.  We added 
modeled Poisson and dark noise to the image, to mimic a modeled PUNCH detector; then 
coded and decoded the noisy gradient image as in Figure \ref{fig:diffmaps_original}.

Figure~\ref{fig:gradient} is a difference image between the original clean gradient 
and the noisy, processed version.  Vertical 
dark artifacts in the processed
image reveal a value-dependent systematic error in the
square-root-processed data.  Figure \ref{fig:first-order-correction} plots the column-average 
difference between the original and decoded gradient images. The systematic offsets have an RMS amplitude
of 0.24 DN; they are therefore strongly statistically significant
toward the lower end of the gradient (left side of plot), and remain  
significant at the upper end of the range we considered.  This is true even though the systematic 
offsets are small compared to the photometric noise or (smaller) square-root coding
error in any one pixel; the imposed coding errors are only revealed upon deep averaging to 
reduce noise and determine the mean value of a large ensemble of pixels\new{ -- in this case, $10^5$
pixels for each integer value between 0-8192}.

We attribute the systematic pattern in the offsets to the fact that the square root operation is
performed with both an integer domain and integer range.  This means that the interval of
photometric
values supported by any one coded value varies slightly from the analytic ideal, yielding local
nonlinear features in the square root mapping.  These variations are larger than the overall
nonlinearity of the analytic square root function, \new{and have a correspondingly larger effect on the mean
of an applied normal distribution of noise.}

\section{\label{sec:better}A Bayesian Square Root Decoder}

To produce a better square root decoder than directly squaring coded values, we used a 
Bayesian
inversion to determine better decoded values.

Each possible coded value \new{$i$} corresponds to a range of ideal photometric input brightnesses. 
Taking $B_i$ to be the na\"ive decoded value \new{corresponding to $c=i$}, 
one may calculate the probability
distribution of values for
a noisy measurement $\hat B_i$ of a pixel with ideal (noise free value) $P$ \newer{that happens to be equal to} $B_i$.  
In the presence of noise, the actual pixel value will be 
\begin{equation}
    \hat P \equiv P + N(\sigma_i) \label{eq:noise}
\end{equation}
where $N(\sigma_i)$ is a normally distributed random variable with sigma corresponding to the 
pixel value $B_i$.  The normal distribution has mean zero, so that the expected value of $\hat P$, 
given the noise-free value of $P$, is just 
\begin{equation}
    \left\{\,\left<\hat P\right>\,|\,(P=B_i)\,\right\} = P = B_i \label{eq:pi-expected-value}\,.
\end{equation}
\new{However, what's available after coding/decoding is not $\hat P$ but $\hat B$. For each $B_i$ value (corresponding 
to coded value $i$), we therefore define a $P_i\equiv B_i$, apply noise to determine $\hat P_i$, and consider the noisy 
value $\hat B_i$ associated with each value $\hat P_i$:}
\begin{equation}
    \hat B_i \equiv D\left(C\left(\hat P_i\right)\right)\label{eq:barBi}\,,
\end{equation}
where $C()$ represents coding and $D()$ represents 
decoding \new{(in this case via Equations~\ref{eq:coding} and \ref{eq:decoding}, respectively)}.  The problem illustrated
in Figure \ref{fig:gradient} is that decoding and coding are not well behaved, so that 
\begin{equation}
    \Delta B_i \equiv \left\{\,\left<\hat B_i\right>\,|\,(P=B_i)\,\right\} - B_i \neq 0\,.
    \label{eq:deltaB}
\end{equation}

$\Delta B_i$ is the systematic offset of the expected value $\left<\hat B_i\right>$ of the 
decoded pixel value $\hat B_i$, given an a priori ``ideal'' measurement
of value $B_i$, the coding scheme, and an a priori known noise distribution. 

One may use $\Delta B_i$ to estimate, via the Bayesian inversion, \new{the expected value 
$\left<P_i\right>$ of the noise-free photometric brightness, given a measured brightness value $\hat B_i$ 
produced by coding and then decoding a noisy photometric value.}  With the assumption
that both the 
noise characteristics and $\Delta B_i$ vary slowly with respect to the coded-value 
index $i$ (as observed in Figure~\ref{fig:gradient}), we can immediately 
write \new{a first-order approximation:}
\begin{equation}
    \left\{\,\left<P\right>\,|\,(\hat B=B_i)\,\right\} \approx B_i - \Delta B_i\,,
    \label{eq:inverted}
\end{equation}
where again $B_i$ is the na\"ively decoded
value corresponding to $c=i$. \new{Equation \ref{eq:inverted} follows from Equation \ref{eq:deltaB}
using Bayes' Theorem \citep[e.g.,][p. 124]{Feller1968} to reverse the roles of measured
and inferred values\newer{, as detailed in an Appendix to this article}.}

\new{The assumption that $\Delta B_i$ varies slowly with respect to $i$ \newer{(or, equivalently, $P$)} is robust over \newer{small changes in $P$ between values of $i$}, 
provided that the coded bit count is set correctly for the a priori known noise width $\sigma_i$. 
That is because $\left<\hat P_i\right>$ is an ensemble average across the entire noise distribution
$N(\sigma_i)$, centered on the interval represented by the coded value $i$.  In the example case 
shown in Figure \ref{fig:coding}, $\sigma_i$ spans an interval of $\pm 4$ coded values over nearly all 
of the original 16-bit dynamic range; this averaging attenuates variations at the $\Delta i=1$ scale by a
factor of order 10, and smaller variations (e.g., between adjacent P values) by a factor that grows as the 
inverse square of the scale.}

With modern computers it is no challenge to assemble a complete decoding table, explicitly 
calculating $\Delta B_i$ for every possible value of $i$\new{, for a complete code table as many
as 20 bits deep.}
We carried out the calculation explicitly for each possible coded pixel value for a 10-bit-deep 
square root
coding scheme operating on 16~bit data, using the modeled
noise levels and camera performance used in Figure~\ref{fig:gradient}.  These camera
effects include modeled dark-current and digitizer offset values, to 
simulate the actual noise
performance of a typical CCD camera; but these features of the noise model are 
negligible compared to 
Poisson noise over most of the coded dynamic range.  
For each \new{$i$} we calculated $B_i$ and enumerated the
probability distribution for $\hat P_i$ across \new{$\pm 4\sigma$} of the modeled Poisson-plus-dark 
noise for the PUNCH detectors, sampling the distribution at \old{5,000}\new{10,000} individual points.  We then 
coded and decoded each of the \old{5,000}\new{10,000} resulting $\hat P_i$ values and \new{explicitly evaluated 
$\left<\hat P_i\right>$ to determine $\Delta B_i$ for each value of $i$ and thereby construct a first-order 
corrected decoding table.}

\begin{figure*}
    \centering
    \includegraphics[width=6in]{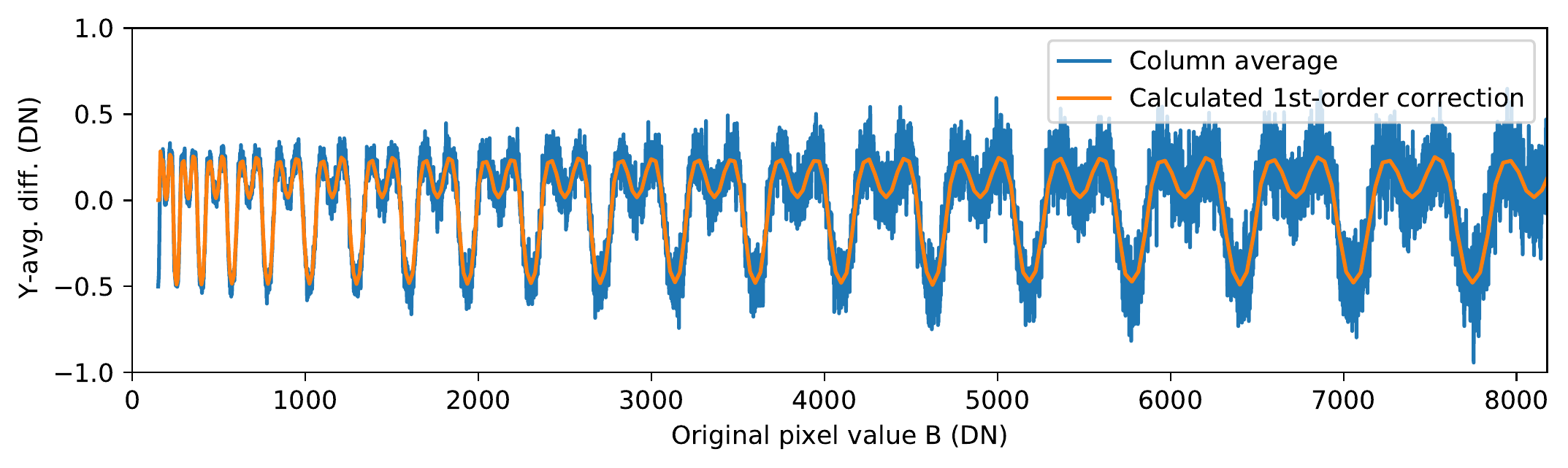}
    \caption{\new{Column-averaged value of all 100,000 rows of the image in Figure \ref{fig:gradient} reveals 
     systematic error from na\"ive square root coding of each possible pixel value, in the presence of noise.  
     First-order correction via Equation \ref{eq:inverted} visually matches the offsets.}}
    \label{fig:first-order-correction}
\end{figure*}

\begin{figure*}
    \centering
    \includegraphics[width=6in]{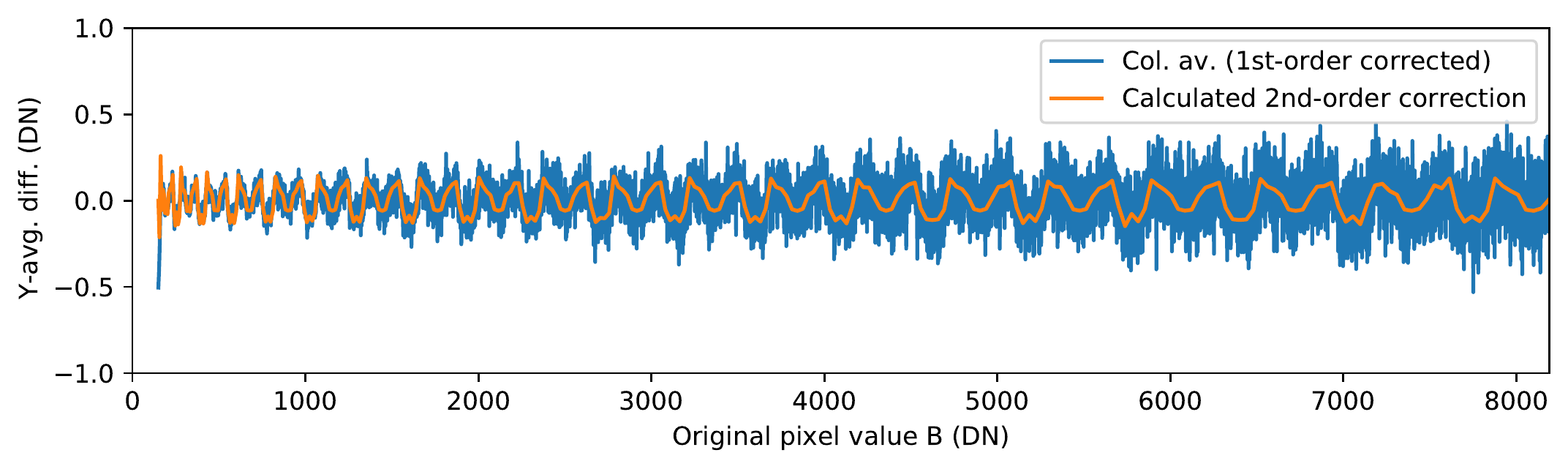}
    \caption{\new{Column-averaged values of the image in Figure \ref{fig:gradient}, with the first-order correction applied, reveal greatly reduced systematic error. A second-order correction, allowing for the rate of change of $\Delta B_i$ with respect to $i$, 
    visually matches the residuals from first-order correction.}}
    \label{fig:grad2}
\end{figure*}

We used lookup into the corrected decoding table 
as a \new{first-order} Bayesian square-root decoder.  Applying that Bayesian decoder to 
the same data as used in Figure~\ref{fig:gradient} reduced the systematic offsets \new{by approximately a factor of four, as shown in Figure~\ref{fig:grad2}; residual systematic fluctuations in the column averages 
in Figure~\ref{fig:grad2} are below, but comparable to, the reduced noise floor after averaging.}  

\new{The blue difference curve in Figure~\ref{fig:grad2} retains obvious low-frequency residuals that follow the
slope of the first-order correction.  Fortunately, second-order correction is also available.
Equation~\ref{eq:inverted} relies on the constancy of the slope $d(\Delta B_i)/d(P_i)$; including 
\newer{the second-order term of the Taylor expansion \new{(based on observing Figure~\ref{fig:grad2})}} yields
}
\begin{equation}
    \new{
    \left\{\,\left<\hat P_i\right>\,|\,(c=c_i)\,\right\} \approx B_i - \Delta B_i - \frac{1}{2}\frac{d\Delta B_i}{di}\,,
    }
    \label{eq:inverted-2}
\end{equation}
\new{where the derivative nomenclature is used to emphasize that $\Delta B_i$ and its rate of change are assumed 
to vary slowly compared to the (integer) steps in $i$.  The orange curve in Figure~\ref{fig:grad2} shows 
this second-order term, calculated from the discrete numeric derivative of the $\Delta B_i$ curve in 
Figure~\ref{fig:gradient}.   
}

\begin{figure*}
    \centering
    \includegraphics[width=6in]{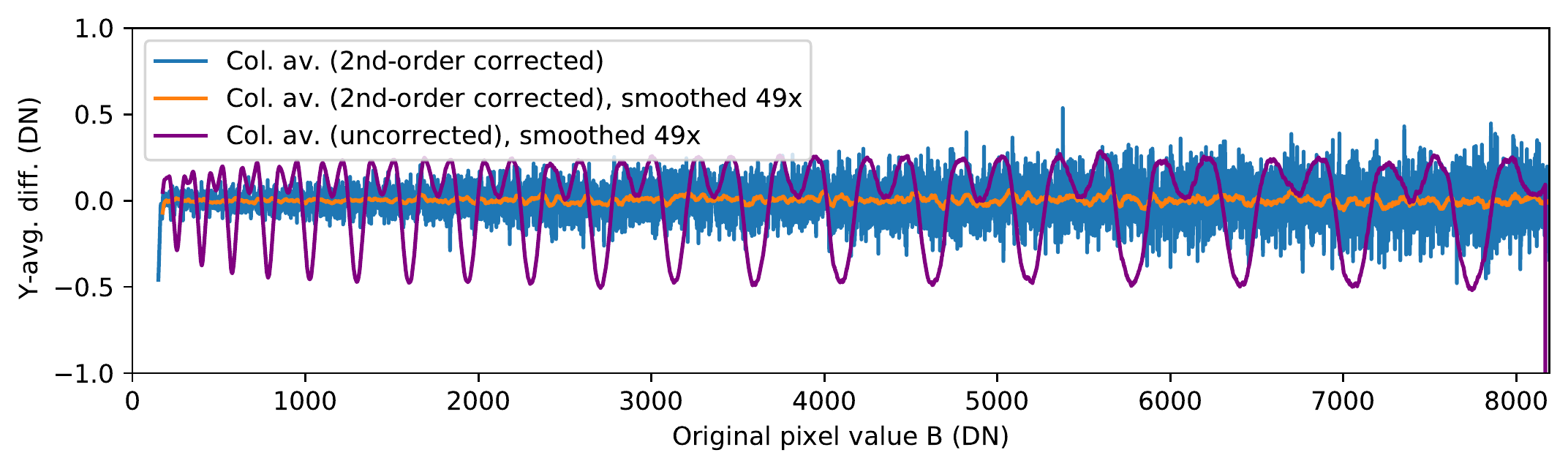}
    \caption{\new{Column-averaged values of the image in Figure \ref{fig:gradient}, with the second-order correction
    applied, reveal still further reduced systematic error. The blue curve is a direct column average across
    all 100,000 samples per value; the orange curve is further averaged across 49 pixel values, beating down 
    noise still further to reveal the systematic offsets.  The uncorrected
    systematic error is also plotted, for comparison.  The 2nd order corrected systematic decoding error 
    is reduced roughly 
    $20\times$ compared to the systematic decoding error from the na\"ive decoding scheme.}}
    \label{fig:grad3}
\end{figure*}

\new{We used Equation~\ref{eq:inverted-2} to assemble a second-order correction table and square root decoder.
Figure~\ref{fig:grad3} shows the result of second-order decoding.  The overlain curves show the 49-value
running mean of the offset, further beating down the noise intrinsic to the image and revealing the systematic
variation.  We found the RMS value of the uncorrected systematic curve, between CCD pixel values of 500 and 4000,
to be 0.23 DN.  The RMS value of the second-order corrected systematic curve over the same range is 0.012 DN,
corresponding to a $20\times$ improvement in systematics; this enables meaningful averaging of up to roughly
$10^6$-$10^7$ 
CCD readout values over this portion of the dynamic range, provided that other systematics are eliminated to
the same degree of precision.}

\new{To test how robust the Bayesian offset estimation might be against errors in the noise model itself, we
regenerated the analysis of Figures \ref{fig:first-order-correction}-\ref{fig:grad3}, using noise models
a factor of two higher or lower than the actual simulated CCD noise level, to identify how sensitive the 
correction might be to exact determination of the noise level in the original measurements.  Shrinking or
expanding the variance of the table
noise model by a factor of two yielded RMS residuals of 0.014 DN and 0.013 DN respectively over the 
500-4000 DN domain. We conclude that the inversion method is robust against factor-of-two discrepancies between
the specific noise model used to calculate the correction table, and the corresponding noise characteristics 
of the original measurement, provided that the dithering condition holds for both the actual and modeled noise level.}

\new{\begin{figure*}
    \centering
    \includegraphics[width=7in]{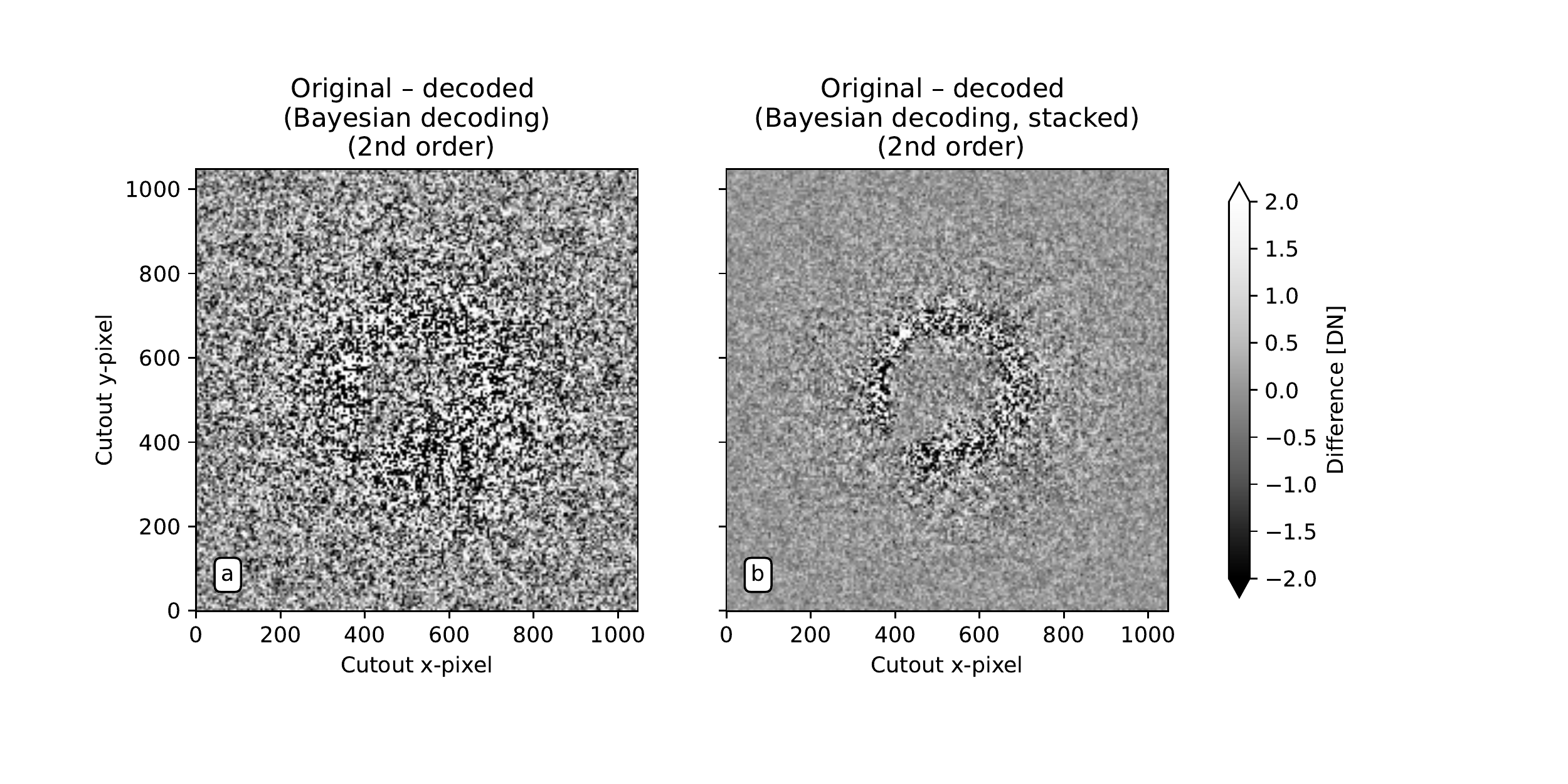}
    \caption{\new{Copy of Figure \ref{fig:diffmaps_original} panels (c) and (d), using the second-order Bayesian decoder instead of direct square-root decoding, eliminates the 
    visual artifacts seen with direct analytic square-root decoding..}}
     \label{fig:diffmaps_bayesian}
\end{figure*}}

\new{To return to the initial dataset, Figure~\ref{fig:diffmaps_bayesian} provides a comparison similar \new{to Figure~\ref{fig:diffmaps_original}}, but 
using the second order Bayesian decoder. In both the difference map in panel (a) and the stacked 
version in panel (b), the visible banding artifacts have been visually eliminated.}

\section{\label{sec:DandC} Discussion \& Conclusions}

We have demonstrated an improved method for square-root decoding that greatly reduces
systematic coding error compared to direct methods. \new{The algorithm is implemented in a Jupyter
notebook that has been released via Zenodo \citep{jupyter_notebook}.}

The improved Bayesian square root decoder we describe is immediately applicable to the
PUNCH mission, which requires averaging over many individual pixels in space and time to 
achieve the photometric precision required by that mission; but it is also applicable to
other instruments and other post-processing schemes that require high photometric precision
relative to the photometric noise in a single measurement.  Square root coding is a 
common technique in data-volume-constrained applications with photon-limited imaging;
Bayesian analysis is important for this and other coding schemes, to ensure that 
photometry is fully preserved in the resultant reconstituted images.  
    
The method we described uses only post-facto analysis to produce a new decoding
table without reference to the actual data being produced, and \new{only moderate (factor-of-two) 
agreement is required between the actual noise characteristics of the instrument and the noise model
used to assemble the table.}  Consequently, the method is applicable not only to new instruments
but also to existing missions or observatories and even to archival data from instruments
that no longer exist, provided that the coding scheme is known.  \new{Further, because the method
uses direct numerical inversion, rather than analytic calculation, to determine corrected decoding 
values, it is applicable to other nonlinear coding schemes than the scaled square root, provided
only that the $\Delta B_i$ from noise-affected ensembles of measurements varies slowly with respect
to the ideal noise-free pixel value P or, equivalently, the coded-value index $i$. In this respect,
our method is more general, though potentially less precise under ideal conditions, than direct 
analytic treatments \citep[e.g.,][]{Bernstein_etal2010} optimized to a particular observing scenario and coding scheme.}

In our particular case, the important nonlinearity
did not arise from the square root operator itself, but rather from the interaction between
the integers and the locations, in the decoded representation, of digital transitions in
the coded values.  In other cases, such as the polarimetric measurements explored by
\citet{Inhester_etal2021}, a mathematical transformation itself, even without integer
transitions, may produce counter-intuitive behavior in the presence of normally distributed
noise.  In both our case and the study by Inhester et al., the most important effect is
that a zero-mean noise distribution can
produce nonzero-mean distributions in transformed data products.  Other higher-order
effects on probability distributions exist, and could in principle cause confusion between,
for example, significant and insignificant features in a noisy data set.

More generally, careful thought is necessary, in order to preserve
the greatest possible utility of the data, when making compromises in representation or
compression.  While first-order effects (such as direct digital transition errors) are
easy to mitigate (for example through dithering), careful thought in advance of 
implementation can also mitigate or remove second order effects (such as the coding 
error we identified here), greatly enhancing the data's utility.

Even more broadly, nonlinear systems interact with noise in counter-intuitive ways. 
Bayesian analysis is a valuable tool to understand the interaction between
nonlinear analysis steps and noise properties of any data product (including reconstituted
``raw'' data), even if the noise sources in the original data are well-behaved.  

\begin{acknowledgments}
This work was funded through PUNCH, a NASA Small Explorer
mission, via NASA Contract No. 80GSFC18C0014.  \new{The work was greatly improved by 
comments from the anonymous referee.}
\end{acknowledgments}

\begin{widetext}
\newer{
\section{Appendix}

In stepping from Equation \ref{eq:deltaB} to Equation \ref{eq:inverted}
we breezily invoked Bayes' Theorem.  Here we demonstrate through analysis how 
that step follows from the theorem.  The inference is 
ultimately tested, and even refined, in the main text 
through numerical treatment rather than direct analysis (e.g., Figures
\ref{fig:grad2} and
\ref{fig:grad3}); but it is a useful exercise to trace the logic back to the 
general theorem.

Bayes' Theorem converts between conditional probabilities.  It is just
\begin{equation}
    \mathbb{P}(A|B) = \mathbb{P}(B|A)\frac{\mathbb{P}(A)}{\mathbb{P}(B)}\,,
    \label{eq:bayes-theorem}
\end{equation}
where $\mathbb{P}$ represents probability of an outcome.  In this case,
the two variables of interest are the \new{true (noise-free)} pixel value $P$ and 
the coded/decoded, noisy pixel value ${\hat B}$.  $P$ has a 
uniform prior distribution; and in the absence of information about $P$, so does
$\hat{B}$ (neglecting edge effects at the top and bottom of the dynamic
range \new{and the very small effect of noise level variation across the width of 
the noise distribution itself}), although the two quantities are related by the normal random 
variable $N$. 
Therefore\new{, to very good approximation},
\begin{equation}
    \mathbb{P}\left(P=j|{\hat B}=B_i\right) =\  
    N_i^{-1}\,\mathbb{P}\left({\hat B}=B_i|P=j\right)\,,
    \label{eq:equivalent}
\end{equation}
\new{where $N_i$ is the number of discrete values of $j$ that convert to the coded value $i$}.
Multiplying each term of Equation \ref{eq:equivalent} by $j$ and summing over all values of $j$ yields:
\begin{equation}
    \left\{\left<P\right>|({\hat B}=B_i)\right\} 
    =\ 
    \sum_j\left\{\frac{j}{N_i} \mathbb{N}\left((j-B_i)/\sigma(B_i)\right)\right\} \approx \sum_k\left\{B_k\frac{N_k}{N_i}\mathbb{N}((B_k-B_i)/\sigma(B_i))\right\}\,,
    \label{eq:expected-value-sumj}
\end{equation}
where $k$ runs over the same values as $i$, and $N_k$ is the number of discrete
values of $j$ that convert to the coded value $k$. The 
second step involves grouping all values of $j$ that lead to a single
coded value $k$. The approximation arises from ignoring variation of $\mathbb{N}$ across those values, which is equivalent to treating $\mathbb{N}$ as piecewise linear across each coded interval.
That egregious sin sidesteps integrating the Gaussian for this approximate analysis.
Extracting $B_i$ from the sum gives
\begin{equation}
    \left\{\left<P\right>|({\hat B}=B_i)\right\}
    \approx
    B_i - \sum_k\left\{B_k(\frac{N_i-N_k}{N_i})\mathbb{N}((B_k-B_i)/\sigma(B_i))\right\}\,,
    \label{eq:final}
\end{equation}
which matches the form of Equation \ref{eq:inverted}. In particular, the
summation term in Equation \ref{eq:final} is the $\Delta B_i$ described in
Equation
\ref{eq:deltaB}; and the $(N_i-N_k)$ coefficient inside the sum embodies 
the discrete-step perturbation described in the main text. 

Contrariwise, choosing $j=B_k$ for some $k$, multiplying by $N_kB_i$, and summing over all values of $i$ yields:
\begin{equation}
    \left\{\left<{\hat B}\right>|P=B_k\right\} \approx \sum_i\left\{\frac{N_k}{N_i}B_i\mathbb{N}\left((B_i-B_k)/\sigma(B_k)\right)\right\}\,,
    \label{eq:penultimate}
\end{equation}
where the equation is only approximate because the same linearization 
of $\mathbb{N}$ is used
as in Equation \ref{eq:expected-value-sumj}.  Extracting $B_k$ from the
sum as above, and then swapping the labels of the $i$ and $k$ indices, yields
\begin{equation}
    \left\{\left<{\hat B}\right>|P=B_i\right\} \approx
    B_i  + \sum_k\left\{B_k(\frac{N_i-N_k}{N_i})\mathbb{N}\left((B_k-B_i)/\sigma(B_i)\right)\right\}\,,
    \label{eq:really-final}
\end{equation}
which matches the form of Equation \ref{eq:deltaB}.  The difference between
Equation \ref{eq:final} and \ref{eq:really-final} arises because the sums are
being carried out over different quantities: Equation \ref{eq:final} is summed over $j$ while Equation
\ref{eq:really-final} is summed
(with weighting) over $i$, although both arise from 
Equation \ref{eq:equivalent}.  The step from Equation \ref{eq:deltaB} to
\ref{eq:inverted}
thus follows analytically from Bayes' Theorem and the approximation given; 
this supports 
the intuition that, to linear order, small-offset perturbations can be 
inverted by merely reversing the direction of the perturbation.  
The piecewise-linear approximation 
used here is somewhat aggressive, and highlights the importance of higher
order correction (described and carried out numerically in the main text).
}

\end{widetext}

\bibliography{noise}{}
\bibliographystyle{aasjournal}

\end{document}